# The Roles of Latent Heating and Dust in the Structure and Variability of the Northern Martian Polar Vortex

E. R. Ball[1] , D. M. Mitchell[1] , W. J. M. Seviour[2,3] , S. I. Thomson[2] , and G. K. Vallis[2]

[1] Cabot Institute for the Environment, and School of Geographical Sciences, University of Bristol, UK; emily.ball@bristol.ac.uk
[2] College of Engineering, Mathematics and Physical Sciences, University of Exeter, UK
[3] Global Systems Institute, University of Exeter, UK

## Abstract

The winter polar vortices on Mars are annular in their potential vorticity (PV) structure, a phenomenon identified in observations, reanalysis, and numerical simulations. Some recent modeling studies have proposed that condensation of atmospheric carbon dioxide at the winter pole is a contributing factor to maintaining the annulus through the release of latent heat. Dust and topographic forcing are also known to be causes of internal and interannual variability in the polar vortices. However, coupling between these factors remains uncertain, and previous studies of their impact on vortex structure and variability have been largely limited to a single Martian global climate model (MGCM). Here, by further decomposing a new MGCM, we decompose the relative roles of latent heat, topography, and dust as drivers for the variability and structure of the northern Martian polar vortex. Additionally, we analyze a reanalysis data set, finding that there are significant differences in vortex morphology and variability according to the spacecraft instrument used for the data assimilation. In both model and reanalysis, high atmospheric dust loading (such as that seen during a global dust storm) can disrupt the vortex, cause the destruction of PV in the low-to-mid-altitudes ($>0.1$ hPa), and significantly reduce spatial and temporal vortex variability. Through our simulations, we find that the combination of dust and topography primarily drives the eddy activity throughout the Martian year, and that although latent heat release can produce an annular vortex, it has a relatively minor effect on vortex variability.

*Unified Astronomy Thesaurus concepts:* Mars (1007); Planetary atmospheres (1244)

## 1. Introduction

Of all the planets in the solar system and beyond, Mars' atmosphere is the best-observed besides Earth and so is one of the best suited for study and comparison with Earth's. On both planets, there are regions of strong mid- to high-latitude zonal winds in the winter hemisphere, known as the polar vortices. While Mars' tropospheric and Earth's stratospheric polar vortices are comparable in their latitudinal extent (the Martian polar vortices extend to around 70°N/S and Earth's to around 60°N/S), the Martian winter tropospheric polar vortices have been shown to be annular in nature, with a local minimum in Ertel's potential vorticity (PV) near the pole (Mitchell et al. 2015). This annular structure, which may be thought of as a ring of high PV enclosed between opposing PV gradients, contrasts with the polar structure of the vortices on Earth, where PV increases monotonically toward each pole, although an annular structure has been found in the mesosphere (Harvey et al. 2009). The annular nature of the Martian polar vortices is likely to affect meridional gas and aerosol mixing, along with vertical wave propagation (Toigo et al. 2017). In general, air in the polar low altitudes is some of the "oldest" (i.e., most isolated) in Mars' atmosphere, but it has been proposed recently that the annulus of PV at the solstice reduces the age of the air in the polar low altitudes through increased mixing with mid-latitude air (Waugh et al. 2019). Mars' polar vortices have been found to extend through the troposphere, decrease in area with height, and retain the same orientation in the vertical (Mitchell et al. 2015).

Unlike on Earth, zonal winds are maximal equatorward of the maximum PV gradient (e.g., Seviour et al. 2017).

Given that a ring of high PV is barotropically unstable (Dritschel & Polvani 1992), the persistence of the annular polar vortices on Mars in observations, reanalyses, and simulations (e.g., Barnes & Haberle 1996; Banfield et al. 2004; Waugh et al. 2016) suggests that there must exist some restoring force that maintains the annulus; Mitchell et al. 2015). Multiple processes have been identified and shown to maintain and stabilize the annular vortex, including diabatic heating by the descending branch of the Hadley circulation (Scott et al. 2020). Recent modeling using a single-layer shallow-water model with a representation of carbon dioxide ($CO_2$) condensation showed that Mars' short radiative timescale may be responsible for stabilizing the annulus (Seviour et al. 2017).

The main constituent of the Martian atmosphere is $CO_2$, which makes up ∼95% of the atmosphere by volume. The $CO_2$ is present in gaseous form and as $CO_2$ ice clouds. In polar regions during the winter seasons, temperatures can fall below the pressure-dependent sublimation point of $CO_2$ (around 149 K), leading atmospheric $CO_2$ to condense and form a layer of $CO_2$ ice on top of the permanent polar ice caps. Latent heat is released into the atmosphere during the phase change from gaseous to solid $CO_2$, and this can increase temperatures in the polar lower altitudes by up to 10 K (Toigo et al. 2017). This means that Mars' atmosphere has a nondilute condensible component, unlike Earth, whose primary condensible component is water vapor, which reaches a molar concentration of up to a few percent. Toigo et al. (2017) showed that in a comprehensive Mars global climate model (MGCM), an annular polar vortex is maintained if the release of latent heat from $CO_2$ condensation is well represented in the model, and that without this forcing, a monopolar vortex (i.e.,

PV increasing monotonically to the pole) forms. Rostami et al. (2018) found that while the annulus is smooth on a timescale of multiple sols (Martian days), the patches of high PV observed when viewing the vortex at a single moment in time are likely caused by the inhomogeneous deposition of $CO_2$.

Regional and global dust storms (GDSs) remain a major influence on Martian atmospheric dynamics, including the polar vortices. Such events provide a major source of interannual variability—one that is not yet fully understood, given the relatively few years of observations. Historically, GDSs appear to dominate the Martian atmosphere approximately once every three southern summers (northern winters; Shirley 2015); there have been only three GDSs since Martian year (MY) 24, occurring in MY 25, MY 28, and, most recently, MY 34.[4] Understanding of the drivers of GDSs remains incomplete, although recent work suggests the influence of solar system dynamics (Shirley 2015) and orbit–spin coupling (Shirley et al. 2020). Regional dust storms and GDSs may partially or fully disrupt the winter polar vortex in what is termed a "rapid polar warming" event, when temperatures rise rapidly within the vortex, apparently a response to increased dust aerosol heating enhancing the meridional circulation (Mitchell et al. 2015; Guzewich et al. 2016). The most recent GDS, occurring in MY 34, is thought to have expanded from an initial equatorial regional storm that created a zonal temperature gradient and hence increased winds, creating a positive feedback (Bertrand et al. 2020). The impact of the MY 34 GDS has been shown to be different in the north and south polar vortices, an effect likely influenced by the timing of the storm, which occurred when the southern polar vortex was decaying (Streeter et al. 2021). The southern polar vortex weakened significantly, while the northern polar vortex remained a robust transport barrier.

In MY 28, a GDS developed at around $L_s \sim 265°$, shortly before the northern solstice. Dust concentrations were elevated (with dust opacity larger than 1) primarily in the midlatitudes but reached up to 40°N (Wolkenberg et al. 2020). The impact of these recent dust storms on the polar vortices has not yet been fully explored, although Guzewich et al. (2016) investigated the impacts of different timings and magnitudes of the GDSs.

Along with atmospheric dust loading, Martian topography may also play a role in the morphology of the polar vortices. The southern hemisphere is strongly dominated by wavenumber 1 waves, but in the northern hemisphere, eddies show a wavenumber 2 pattern (Hollingsworth & Barnes 1996), likely influenced by the hemispheric topographical asymmetries. It is also thought that the northern polar vortex has an elliptical shape, on average, in part due to topographically forced zonal wavenumber 2 waves (Mitchell et al. 2015; Rostami 2018). Indeed, recent work has shown that the suppression of wavenumber 2 stationary waves during the MY 34 GDS corresponded to the reduction of the ellipticity of the northern polar vortex (Streeter et al. 2021). In both winter hemispheres, there is a "solsticial pause" in the amplitude of low-altitude transient waves. Studies have attributed this in part to topographic zonal asymmetry (Lewis et al. 2016; Mulholland et al. 2016).

It is not yet fully understood how the Martian polar vortices are influenced by the interplay of topography, latent heating, and dust loading. Guzewich et al. (2016) and Toigo et al. (2017) investigated how dust and latent heat release each separately influence the polar vortices in modeling studies, but there has not yet been any study investigating the combined effects. In this paper, we investigate the transience and variability of the northern Martian polar vortex through the use of a reanalysis data set and idealized simulations from a newly developed Martian configuration of the flexible modeling framework Isca (Vallis et al. 2018; Thomson & Vallis 2019a). We focus here on the northern polar vortex due to both previous work suggesting that the northern hemisphere exhibits a stronger solsticial pause (Lewis et al. 2016) and the findings of Guzewich et al. (2016), who noted that the northern vortex is more heavily influenced by dust loading than the southern vortex. In our simulations, the southern vortex is found to be reasonably invariant to southern winter dust loading.

Thanks to prolonged observations of the Martian atmosphere, there are sufficient data available to create a Martian reanalysis data set spanning several MYs, of which there are currently three available (Montabone et al. 2014; Greybush et al. 2019; Holmes et al. 2020). A reanalysis data set assimilates observations into a general circulation model to provide a three-dimensional, gridded estimate of the atmospheric state, including variables that cannot be directly measured. We use the newly developed Open access to Mars Assimilated Remote Soundings (OpenMARS) reanalysis product (Holmes et al. 2020) to investigate the features of the northern Martian polar vortex. Previous studies of Martian polar vortices in reanalyses have focused particularly on MY 24–27 (Mitchell et al. 2015; Waugh et al. 2016). We are particularly interested here in the impacts of the MY 28 GDS, as this is the only solsticial GDS in the reanalysis period and provides an exciting opportunity to study the effect of dust on the northern polar vortex during this time. The GDSs primarily occur between $L_s \sim 200°$ and $340°$ (Kass et al. 2016), leading to there being no equivalent southern winter solsticial GDS during the reanalysis period.

We aim to understand the mechanisms that drive the northern Martian polar vortex. We identify significant features of the vortex in OpenMARS and explore these using the flexible modeling framework Isca. Using Isca, we perform an attribution-type study of the polar vortex, with topography, latent heating, and dust parameters to be changed systematically. The rest of the paper is outlined as follows. Section 2 introduces the quantities that we will use to investigate the vortex and the reanalysis products currently available and describes the model used in this study. From there, we discuss the northern polar vortex as seen in the reanalysis and our simulations in Section 3. We present results describing the climatological state of the polar vortex and interannual variability in Section 3.1 and the subseasonal variability in Section 3.2. Finally, we summarize the study in Section 4.

## 2. Methods

### 2.1. Potential Vorticity

PV is a dynamically important quantity, the product of absolute vorticity and the gradient of potential temperature, that is particularly useful in the study of polar vortices. In general, PV is materially conserved provided that frictional and diabatic processes vanish. In this paper, we use an approximation to the true PV, valid to a good approximation in a hydrostatic

---

[4] The MYs are numbered according to Clancy et al. (2000), where MY 1 begins at the northern spring equinox ($L_s = 0°$) on 1955 April 11. Each MY is roughly the length of 2 Earth yr, and midwinter in the northern hemisphere is at $L_s = 270°$. Thanks to a recent increase in the number of satellites orbiting Mars, there has been almost continuous observation of the Martian atmosphere since 1999, corresponding to $L_s \sim 104°$, MY 24.

atmosphere, namely,

$$q(\theta, t) = -g(f + \zeta_p)\frac{\partial \theta}{\partial p},\tag{1}$$

where $q$ is PV, $g$ is the Martian gravitational acceleration ($3.72\,\mathrm{m\,s^{-2}}$), $f$ is the Coriolis parameter, $\theta$ is the potential temperature, $\zeta_p$ is the vertical component of relative vorticity evaluated on a pressure surface, and $p$ is the pressure (Hoskins et al. 1985; Read et al. 2007). The PV is calculated from winds and temperature as a function of pressure in the reanalysis and model, then linearly interpolated to isentropic surfaces for our analysis. To remove the large vertical variation of PV in the atmosphere, we then use a common scaling of PV devised by Lait (1994). To be consistent with Waugh et al. (2016), the exact scaling chosen is

$$q_s(\theta, t) = q \times \left(\frac{\theta}{\theta_0}\right)^{-(1+c_p/R)},\tag{2}$$

with $\theta_0 = 200$ K an arbitrary reference potential temperature and $c_p/R = 4.0$ the ratio of specific heat at constant pressure to the specific gas constant of Martian air.[5] A common way to express PV in Earth's atmosphere is by using potential vorticity units (PVU). Here we present PV in Martian PVU (MPVU), where $1\,\mathrm{MPVU} = 100\,\mathrm{PVU} = 10^{-4}\,\mathrm{K\,m^2\,kg^{-1}\,s^{-1}}$, adopting the convention used in Streeter et al. (2021).

### 2.2. Reanalyses

Currently, three reanalysis data sets are available for the Martian atmosphere. These are the Mars Analysis Correction Data Assimilation (MACDA; Montabone et al. 2014), Ensemble Mars Atmosphere Reanalysis System (EMARS; Greybush et al. 2019), and OpenMARS (Holmes et al. 2020) data sets. We primarily use OpenMARS within this work. OpenMARS assimilates observations of thermal profiles; water ice and dust opacities; and ozone and water vapor column abundances into the UK-LMD MGCM to produce a gridded estimate of Martian weather spanning MY 24–32. Full details of the profiles assimilated into OpenMARS and the underlying MGCM may be found in Holmes et al. (2020), although we briefly discuss the relevant details here. The OpenMARS data set can be broadly separated into two distinct periods based on the retrieval instruments used. In the era MY 24–27, temperature retrievals assimilated into the model are from the Thermal Emission Spectrometer (TES) on board the Mars Global Surveyor (MGS). The TES nadir retrievals provide coverage of temperatures up to around 40 km in altitude but have the largest uncertainties at the lowest altitudes due to possible errors in estimating surface pressure. Systematic errors in temperature retrievals peak over the winter polar regions due to cold surface temperatures, and there is a lack of coverage of column dust optical depth (CDOD) retrievals at winter high latitudes. Due to cold surface temperatures on the nightside of the planet, only dayside dust retrievals are assimilated. In MY 28–32, retrievals are from the Mars Climate Sounder (MCS) on board the Mars Reconnaissance Orbiter. The MCS temperature profiles have greater vertical resolution than TES (5 km rather than 10 km) and cover up to approximately 85 km in altitude

(Holmes et al. 2020). Conversely to TES, temperature retrieval errors are lowest in the lower atmosphere for MCS. There are an increased number of MCS profiles at the end of MY 28, in an effort to observe the atmosphere during the MY 28 GDS. Finally, although estimates of CDOD from MCS observations have the possibility of error due to the extrapolation down to the surface, retrievals are possible during both daytime and nighttime polar winter. It is worth noting that there is no overlap in the TES and MCS temperature retrievals, so there are no reanalysis data available for the northern hemisphere winter of MY 27.

OpenMARS may be seen as the updated version of the MACDA data set (the details of which are described in Montabone et al. 2014), which spans MY 24–27 (assimilating TES thermal profiles as in OpenMARS) and is based on an older version of the same MGCM. OpenMARS in the TES period differs from MACDA in that the underlying model has been updated; for example, MACDA uses an analytical dust distribution, whereas in OpenMARS, dust is freely transported, and OpenMARS now includes a thermal plume model. A detailed description of MACDA and a discussion of how the products differ may be found in Montabone et al. (2014) and Holmes et al. (2020), respectively.

The final available reanalysis data set is EMARS, which spans MY 24–34. Full details of the reanalysis may be found in Greybush et al. (2019) and its precursor Greybush et al. (2012). EMARS is a 16-member ensemble reanalysis that uses the Geophysical Fluid Dynamics Laboratory MGCM along with assimilation of retrievals from TES and MCS for the periods described above. Dust is controlled by three radiatively active tracers. Temperature retrievals that fall significantly below the pressure-dependent $CO_2$ condensation point, $T_c$, are modified to match the condensation temperature, and when temperatures within the model are projected to be below $T_c$, gaseous $CO_2$ is removed from the atmosphere and placed on the surface as $CO_2$ snow. In contrast, for OpenMARS, when temperature retrievals fall significantly below $T_c$, they are simply filtered out before assimilation into the model.

We here present results using the OpenMARS data set, although we do verify any results that we find in OpenMARS against EMARS and find broadly similar results. Where there are any notable differences, we include supplementary figures showing the results in EMARS and discuss these differences.

### 2.3. Model

In addition to the OpenMARS reanalysis data set, we use an idealized climate model to investigate the role of different physical processes in shaping polar vortex structure and variability. We make use of Isca, an idealized modeling framework developed with flexibility in mind, first described in Vallis et al. (2018). One significant advantage of Isca, which we exploit here, is the relative ease of including or excluding different physical processes within the model configurations. Here we include a brief description of the Isca representation of the Martian atmosphere we build upon in this work; additional details can be found in Section 4 of Thomson & Vallis (2019a; our simplest "control" configuration is identical to the configuration described in their Section 4.3). Isca uses a spectral, primitive equation dynamical core in spherical coordinates along with the multiband, comprehensive SOCRATES radiation scheme (Manners et al. 2017; Thomson & Vallis 2019b). The spectral files used were originally created for the ROCKE-3D model (Way et al. 2017) and have been adapted to include dust aerosols, as described below. We use a T42 spectral resolution (corresponding to a $64 \times 128$ ($\sim 2.8° \times 2.8°$) spatial grid) and 25 vertical sigma levels reaching approximately 0.05 Pa. We use topography from the Mars Orbiter Laser Altimeter (MOLA) measurements on board MGS (Smith et al. 1999).

---

[5] The code used for calculation of PV and all other analyses in this paper is available in Ball (2021).

To investigate the potential dynamical and thermal impacts on the northern Martian polar vortex, we have additionally developed and implemented[6] an idealized dust scheme and representation of latent heating due to the condensation of $CO_2$, described below. These were identified to be the primary missing processes in the representation of Mars' atmosphere within the Thomson–Vallis configuration of Isca–Mars and, as shown below, are fundamental in attaining a reasonable representation of the polar vortices in the model. We do not include any representation of radiatively active ice clouds, which have been shown to significantly influence temperatures and atmospheric circulation in MGCMs (Madeleine et al. 2012).

### 2.3.1. Representation of Latent Heating from Carbon Dioxide Condensation

For this work, we have developed a simple representation of the latent heat released from $CO_2$ condensation, as this has been proposed to play a crucial role in driving the annular polar vortex in MGCMs (Toigo et al. 2017). The latent heat release from $CO_2$ condensation in Isca does not involve any $CO_2$ phase changes but rather a simple prescribed temperature tendency once the condensation point of $CO_2$ is reached. Following Lewis (2003) and Way et al. (2017), the condensation point of $CO_2$, $T_c$, is derived from an approximate solution to the Clausius–Clapeyron relation, namely,

$$T_c = 149.2 + 6.48 \log(0.00135p), \tag{3}$$

where $p$ is the model pressure in pascals and $T_c$ is in kelvins. When the temperature projected by the model, $T^*$, falls below $T_c$, the model temperature $T$ is set to $T_c$, as in Forget et al. (1998). The difference $T_c - T^*$ is used to calculate the amount of latent heat that would be released by estimating the mass of $CO_2$ that would condense according to this temperature difference.

We have not yet implemented within Isca any representation of the mass loss itself that occurs when $CO_2$ condenses. Although this can be a significant amount—up to 30% of the mass of the atmosphere can be lost over the course of a Martian winter (Tillman 1988)—it has previously been shown that the dynamical effect of the pressure changes caused by the mass loss has less impact on the structure of the northern Martian polar vortex than the latent heat release associated with the $CO_2$ condensation (Waugh et al. 2019). As we represent dust using an analytical profile rather than as a tracer, there is no opportunity for dust particles to act as condensation nuclei for $CO_2$ ice, a process that allows more atmospheric $CO_2$ to condense in a dusty atmosphere (Gooding 1986). There is also no representation of $CO_2$ sublimation from surface ice into the atmosphere. In this regard, our choice not to update the pressure where $CO_2$ condensation occurs makes sense in that otherwise, eventually, the atmospheric mass would disappear if we did consider the mass lost.

### 2.3.2. Dust Scheme

Dust is known to be an important feature in many Martian atmospheric processes. We choose a simple representation of dust—prescribing an analytical latitudinal and vertical profile that has been used in several MGCMs without explicitly modeling dust lifting processes.

The effective radius and variance of dust particles are given by $r_{eff} = 1.5 \, \mu m$ and $\nu_{eff} = 0.3 \, \mu m$, respectively, consistent with simulation 2 of Madeleine et al. (2011). The size distribution of the dust particles follows a modified-gamma profile, and the spatial distribution of dust radiative properties is uniform. The parameters of the modified-gamma distribution are determined by the values of $r_{eff}$ and $\nu_{eff}$ according to Hansen & Travis (1974). The refractive indices of dust have been chosen according to Wolff et al. (2006). Scattering properties are then calculated using a Mie scattering algorithm. Although dust particles have been noted to be cylindrical with a diameter-to-length ratio of 1.0 (see, for example, Wolff et al. 2009), here our choice of the SOCRATES radiation scheme means that they must be modeled as spherical. However, the approximation does not introduce systematic effects above the 5% level in radiance (Wolff et al. 2006) and is computationally much more efficient.

Zonally averaged infrared absorption CDOD normalized to the reference pressure of 610 Pa product is input and converted to a surface mass mixing ratio by the SOCRATES interface within the model. The vertical and longitudinal distribution of dust in the model then follows the modified Conrath $\nu$ profile described in Montmessin et al. (2004). The top of the dust layer is given by $z_{max}$ (kilometers), dependent on solar longitude $L_s$ (degrees) and latitude $\phi$ (degrees), calculated as

$$z_{max}(L_s, \phi) = 60 + 18 \sin(L_s - 158)$$
$$- (32 + 18 \sin(L_s - 158)) \sin^4 \phi$$
$$- 8 \sin(L_s - 158) \sin^5 \phi. \tag{4}$$

From this, the dust mass mixing ratio $a$ is calculated using the following expression:

$$a = a_0 \exp \left\{ \nu \left[ 1 - \max \left\{ \left( \frac{p_{ref}}{p} \right)^{70 \, km/z_{max}}, 1 \right\} \right] \right\}, \tag{5}$$

where $a_0$ is a constant mass mixing ratio at pressure $p_0$, determined by dust opacity at the surface (see Conrath 1975; Pollack et al. 1979, for details), which is scaled to produce realistic model temperatures and winds. Here $p_{ref}$ is the reference pressure of 700 Pa. The Conrath $\nu$ profile allows well-mixed dust in the lower atmosphere with exponentially decreasing concentrations of dust at higher altitudes and has been used in many MGCMs, such as Montmessin et al. (2004), Madeleine et al. (2011), and Guzewich et al. (2016). However, recent work showing evidence of detached dust layers suggests that this profile might not be as appropriate as once thought, particularly in the tropics (Heavens et al. 2014). We choose, however, to use it here for its simplicity, ease of adaptation, and comparability with other models. Since dust is determined by an analytical expression within the model and is not lifted from the surface into the atmosphere, there is no capacity for spontaneous generation of dust storms. However, the flexibility to choose an analytical distribution within the model allows the

---

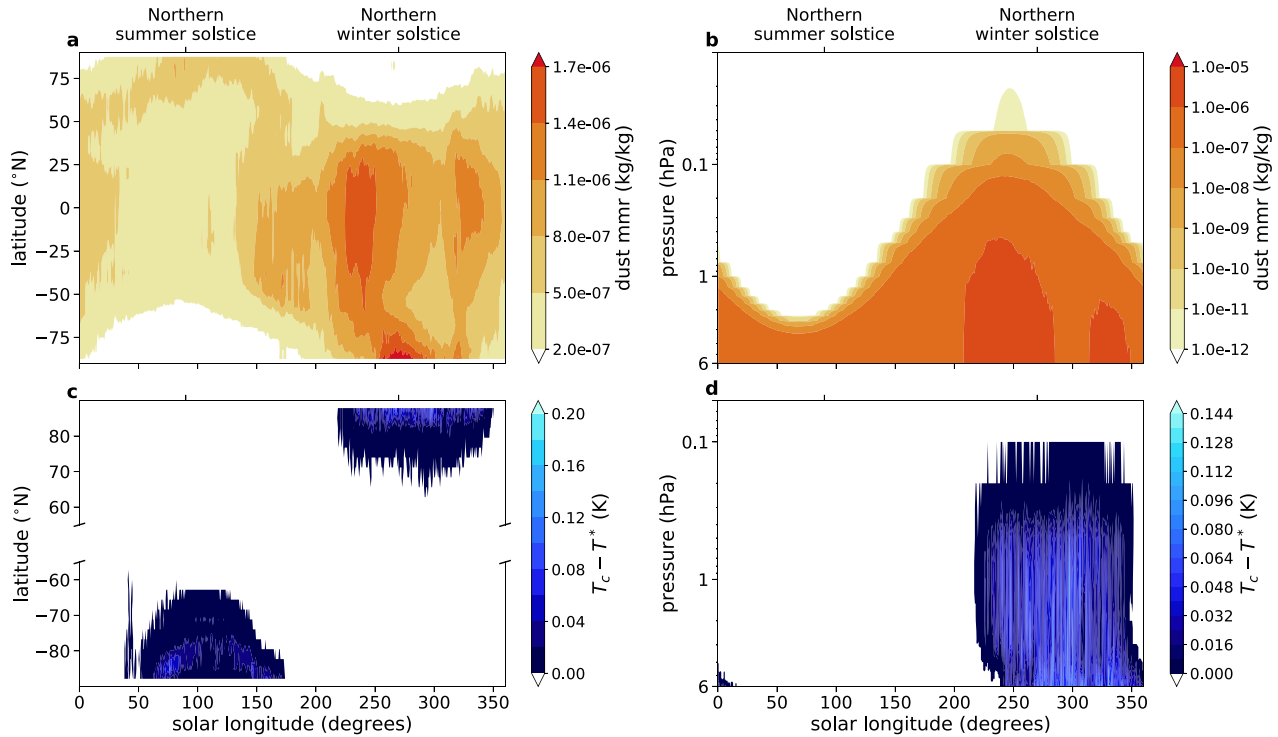

**Figure 1.** Evolution of (a) surface and (b) vertical equatorial dust mass mixing ratios for the "climatological" dust product over the course of an MY. Temperature below condensation point, defined as $T_c - T^*$, in the "Latent Heat and Dust (Topography)" simulation (c) on the 2 hPa surface and (d) at 85 °N.

user to investigate the impact of high dust loading in an area of their choosing. All dust distributions underlying the simulations in this study are zonally symmetric, although Isca presents the user with the option of a full latitudinally and longitudinally varying dust distribution.

### 2.3.3. Dust Scenarios

We present results from simulations from Isca with various dust scenarios. The model uses the Mars Climate Database (MCD; Forget et al. 1999; Millour et al. 2018) dust year climatologies (Montabone et al. 2015, 2020) to form simulations that isolate the role of interannual dust variability, the years of which may be compared directly to the years available in OpenMARS. Details of the dust cycle used in the MCD are described in Madeleine et al. (2011). These dust products inform the surface dust mass mixing ratio, $a_0$, and all years of simulations then follow the Conrath $\nu$ vertical profile (Equation (5)). These "yearly" simulations all include representations of dust, latent heat release, and topography, as we are interested in the influence of dust in each MY in our most realistic simulations.

We also use the MCD standard dust scenario, which is built by averaging the kriged yearly climatologies MY 24–31 (excluding the GDS events of MYs 25 and 28; Montabone et al. 2015), to investigate how latent heat release, dust, and topography influence the polar vortex in a suite of process-attribution simulations. Using this climatological dust product allows us to simulate our best guess of an MY with "typical" dust loading. The evolution of surface and vertical dust mass mixing ratios is shown in Figure 1 for the process-attribution dust simulations. An explicit list of the simulations performed, processes turned on, and dust products used in each can be found in Table 1.

Each simulation was run for 4 MYs, and the results shown are the average of the 3 final yr (ensemble members) of each simulation. The first year is discarded, as this is found to be sufficient to reach an equilibrated state from rest. The exception is MY 28, which was run for 10 yr to fully investigate the impact of the GDS in that year. Given the short radiative timescales on Mars and the lack of freely moving dust or $CO_2$ microphysics, there is little ensemble spread in general. Figure 1 shows the evolution of the surface and equatorial vertical dust mass mixing ratio throughout the "climatological year," illustrating the peak in dust loading around the northern winter solstice. The vertical profile is informed by the Conrath $\nu$ profile. Absorption CDOD normalized to 610 Pa for individual years can be seen in Figure 21 of Montabone et al. (2015), illustrating dust loading across individual MYs.

Figure 1 also illustrates a proxy for the amount of latent heat released over the course of an MY during the "Latent heat and dust (topography)" simulation, including dust, topography, and latent heat release. The proxy for latent heat release, defined as $T_c - T^*$, shows where the temperature $T^*$ falls below the pressure-dependent condensation point of $CO_2$ as given in Equation (3), thus representing both where and how much latent heating is occurring. We see that latent heat release occurs throughout northern winter up to ∼0.1 hPa.

## 3. Results

### 3.1. Polar Vortex Mean State and Interannual Variability

We will initially discuss the structure of the mean-state winter (averaged over $L_s = 270°–300°$) polar vortex. Figure 2 shows the vertical cross section of zonal-mean PV over each winter of OpenMARS data. The PV has been scaled according to Lait (1994) to remove vertical variation due to exponentially decreasing pressure with altitude (see Section 2.1 for details). It

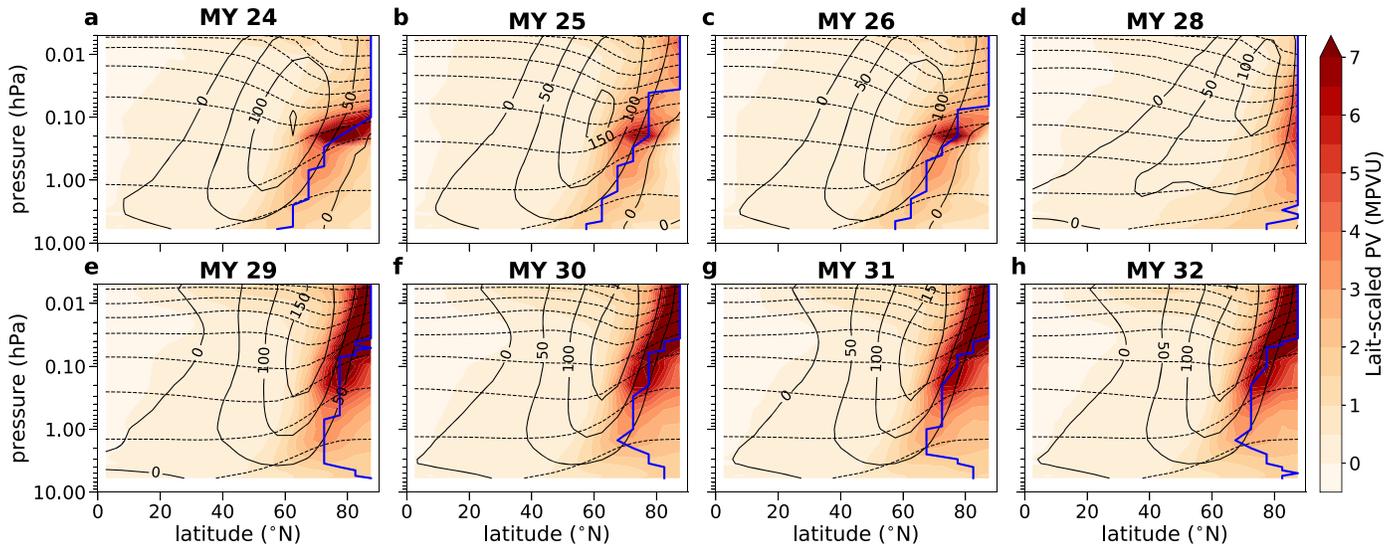

**Figure 2.** Winter-mean, zonal-mean, Lait-scaled PV (shading) and zonal-mean zonal wind (solid black contours) for each Martian winter in the OpenMARS reanalysis data set. Note that the winter of MY 27, when there were no TES or MCS temperature retrievals, is not shown. Dashed contours correspond to the 200, 300, … K potential temperature surfaces, and the solid blue contour marks the latitude at which PV takes its maximum value at each pressure level. The Martian winter solstice falls at $L_s = 270°$, and each panel is averaged over $L_s = 270°–300°$.

**Table 1**
List of Simulations Performed

| Simulation Name | Short Name | Topography | Latent Heating | Dust | Dust Product, Scaling |
|---|---|---|---|---|---|
| Control | Control | None | No | No | … |
| Latent heat | LH | None | Yes | No | … |
| Dust | D | None | No | Yes | Climatology, 1x |
| Latent heat and dust | LH+D | None | Yes | Yes | Climatology, 1x |
| Topography | T | MOLA | No | No | … |
| Latent heat (topography) | LH+T | MOLA | Yes | No | … |
| Dust (topography) | D+T | MOLA | No | Yes | Climatology, 1x |
| Latent heat and dust (topography) | LH+D+T | MOLA | Yes | Yes | Climatology, 1x |
| Yearly | Yearly | MOLA | Yes | Yes | Yearly (MY 24–32), 1x |
| High dust | High dust | MOLA | Yes | Yes | Yearly (MY 24–32), 2x |

**Note.** The scaling of the dust product refers to the scaling of the surface mass mixing ratio, $a_0$, compared to its value in the "Dust" simulation. In all simulations, the dust mass mixing ratio follows the Conrath $\nu$ profile.

can be seen that the maximum in PV (blue contour) lies away from the pole, meaning an annular vortex below ∼0.1 hPa in each MY, save MY 28. In MY 28, the zonal-mean zonal winds are weaker, and the maximum in PV lies at the pole.

We also note that there appears to be a systematic difference in the vertical structure of the PV cross section in the different periods of OpenMARS. The TES period (MY 24–27; Figures 2(a)–(c)) shows a very strong vortex core confined roughly below 0.1 hPa, whereas the MCS period (MY 28–32; Figures 2(d)–(h)) displays a stronger vortex at higher altitudes. EMARS also displays this high PV at high altitudes in the MCS period (not shown). Since MCS temperature retrievals reach approximately 80 km (∼0.05 hPa) in altitude, compared with TES retrievals at ∼40 km (∼3 hPa), this would reinforce the finding of Waugh et al. (2016) that the differences between the MACDA (noting that MACDA is only available for the TES period and shows a very similar structure to OpenMARS TES) and EMARS vortex structure above 0.1 hPa are largely due to differences in the underlying models, and the reanalyses below 0.1 hPa are controlled by the assimilated data. We also note that the observations assimilated in the MCS era lead to a stronger jet core that extends to higher altitudes, with zonal-mean zonal winds exceeding $150 \, m \, s^{-1}$ reaching well into the upper atmosphere (<0.01 hPa).

Figure 3 shows winter Lait-scaled PV and zonal winds in OpenMARS on the 300 K isentropic surface. The annular elliptical vortex is visible in all years, save MY 28 (Figure 3(d)). In MY 28, we also observe reduced zonal wind speeds (by up to $40 \, m \, s^{-1}$) and the destruction of the polar vortex in the winter period. The GDS of MY 25 (Figure 3(b)), which ended at around $L_s \sim 245°$, appears not to have had this effect. Given the relatively short radiative relaxation timescales on Mars (∼0.5–2 sols; Eckermann et al. 2011), this is perhaps not surprising. Indeed, an average over the period $L_s = 235°–245°$ equivalent to Figure 3 shows that, during this period, the polar vortex in MY 25 was greatly weakened but clearly recovers by the winter solstice (not shown).

In order to understand the structures seen above, we sequentially "turn on" certain processes within Isca, thus forming a hierarchy of process-attribution simulations. Figure 4 shows that a representation of dust is necessary within the model to achieve the high zonal winds and temperatures seen in OpenMARS. This is due to the atmospheric heating caused by the addition of dust aerosols into the atmosphere. Dust within the model enhances both the jet strength and the polar high-altitude warming, a characteristic feature of the Martian winter atmosphere caused by enhanced poleward meridional transport of warm air (Guzewich et al. 2016). Polar temperatures at low altitudes (>1 hPa) in simulations that include latent heat release (Figures 4(c) and (e)) are up to 10 K warmer than those without (Figures 4(b) and (d)), consistent with the amount of warming seen in Toigo et al. (2017). This is in much better agreement with the temperatures seen in reanalysis (Figure 4(a)). However, we note that the small area over which $CO_2$ condenses appears insufficient to strongly affect zonal-mean zonal winds, aside from a small weakening in the winds.

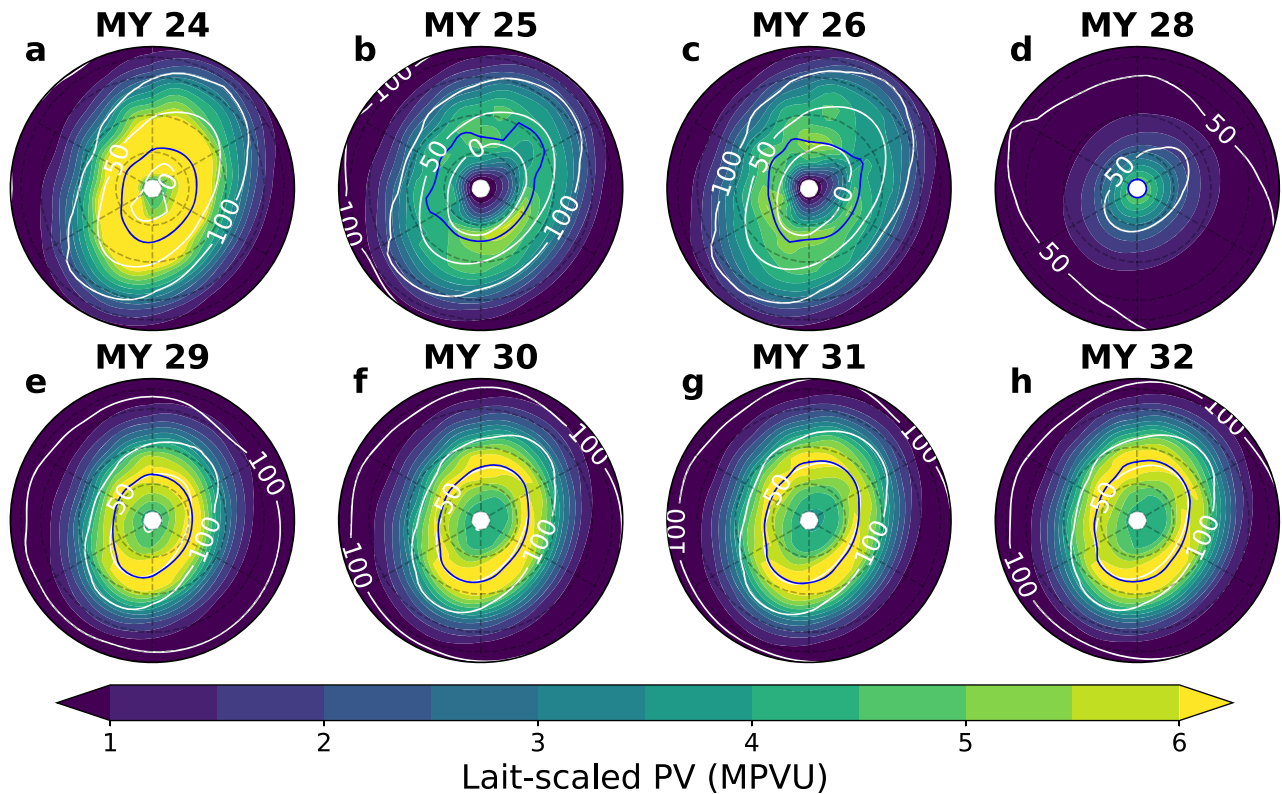

**Figure 3.** Winter-mean ($L_s = 270°–300°$) north polar stereographic maps of Lait-scaled PV (shading) and zonal wind (contours) on the 300 K (~0.5 hPa) surface from each winter of OpenMARS data. The solid blue contour shows the latitude of maximum PV. Dashed latitude lines correspond to 60°N, 70°N, and 80°N, with each panel bounded at 55°N.

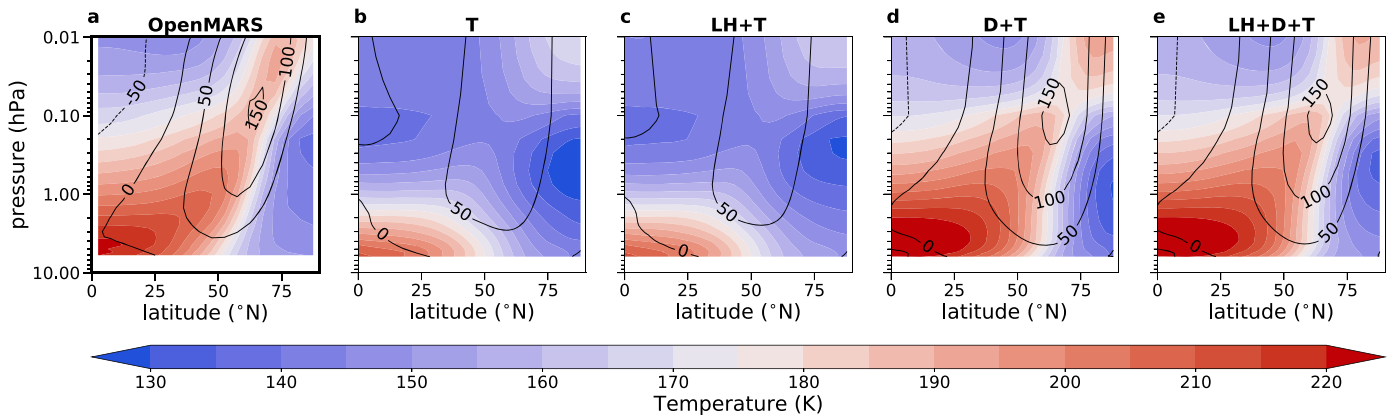

**Figure 4.** Winter cross-sectional profiles ($L_s = 270°–300°$) of zonal-mean temperature (shading) and zonal wind (contours) in the northern hemisphere for (a) all years of OpenMARS and (b)–(e) the process-attribution simulations with topography. For each simulation, winds and temperatures have been averaged over three ensemble members.

Figure 5 shows winter zonal-mean PV for various Isca configurations, all with topography. By observing the solid blue contour in Figures 5(c) and (e), it is clear to see that latent heating in the model does indeed push the maximum in PV away from the pole, particularly at altitudes between 1 and 0.1 hPa, compared to simulations that do not include representation of latent heat release. This suggests that latent heat release at the pole is acting to reduce PV there, thereby driving the annulus. We also see that dust in the model pushes the jet core further poleward, reducing the latitudinal extent of the polar vortex, as may be expected from an enhancement of the mean meridional circulation. Dust also increases the vertical potential temperature gradient at lower altitudes (consider the tightening of the isentropes seen in Figures 2(b) and (d), for example), and it is this that contributes to the strengthening of PV at around 1 hPa.

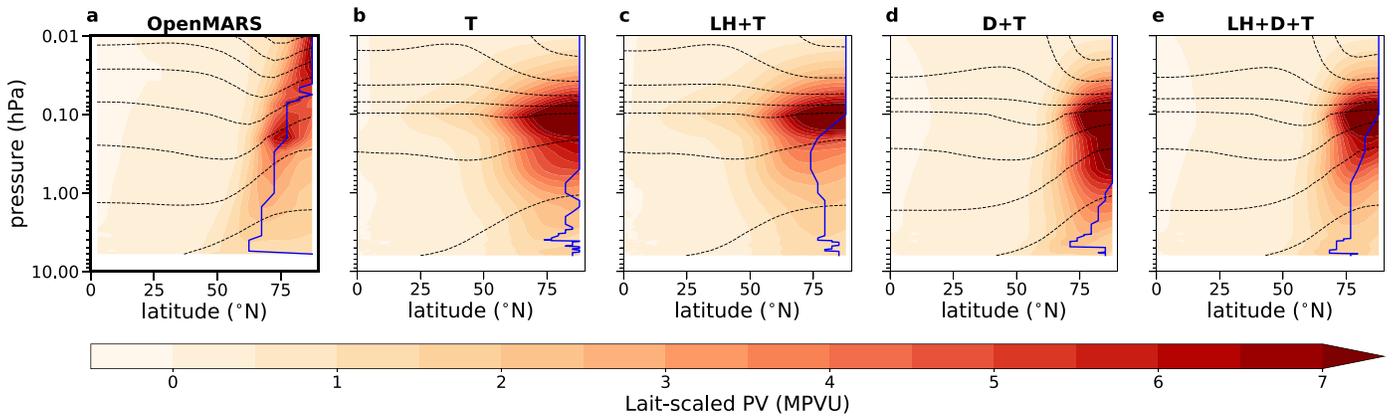

**Figure 5.** Winter ($L_s = 270°$–$300°$) zonal-mean Lait-scaled PV (shading) and potential temperature surfaces (dashed contours, corresponding to 200, 300, …K) in the northern hemisphere for (a) all years of OpenMARS and (b)–(e) the process-attribution simulations with topography. For each simulation, winds and temperatures have been averaged over three ensemble members. The solid blue contour shows the latitude of maximal Lait-scaled PV at each pressure level, as in Figure 2.

Another way to visualize the influence of turning on processes in our process-attribution simulations is to consider polar stereographic projections of the polar vortex, as shown in Figure 6. We note several effects. Turning on latent heating increases the width of the annulus (the distance from the pole to the maximum value in PV) in all experiments. Turning on dust may actually also increase the width of the annulus in those experiments where latent heating is not present (see Figures 6(a) and (e)), likely due to the downwelling of warm air from enhanced meridional circulation. However, when latent heating is also turned on, dust reduces the width of the annulus (e.g., Figures 6(c) and (g)), suggesting a complex relationship between these two factors. Turning on topography always reduces the width of the annulus when one exists (Figures 6(c)–(h)), a result consistent with Seviour et al. (2017), who found that topographic forcing of a large enough amplitude would lead to a monotonic PV distribution in a shallow-water model. Turning on topography also forces the elliptical wave 2 shape of the northern polar vortex. The ellipse is particularly noticeable in simulations without dust (Figures 6(b) and (d)). This is consistent with previous studies that showed that increased loading during the MY 34 dust storm pushed the jet poleward away from the wavenumber 2 topographic forcing, thus reducing the ellipticity of the vortex (Streeter et al. 2021).

To investigate the influence of interannual dust variability, we also ran the yearly simulations described in Table 1. Figure 7 shows PV for each year on the 300 K surface, which is just below the core of the vortex in OpenMARS. Although the dust signal is relatively weak compared to the reanalysis, MY 28 (Figure 7(d)) sees the annulus shrink toward the pole, particularly noticeable when only compared to other MCS years (MY 29–32; Figures 7(e)–(h)). On the 300 K level, PV actually strengthens in MY 28, although the weakening seen in OpenMARS does occur at higher levels (>350 K; not shown). Although the zonal winds do not appear much weaker in the MY 28 simulations, as they are in the reanalysis (Figure 3(d)), when the dust mass mixing ratio is increased further within the model, thereby increasing the total atmospheric dust loading, the zonal winds weaken and PV is completely destroyed in MY

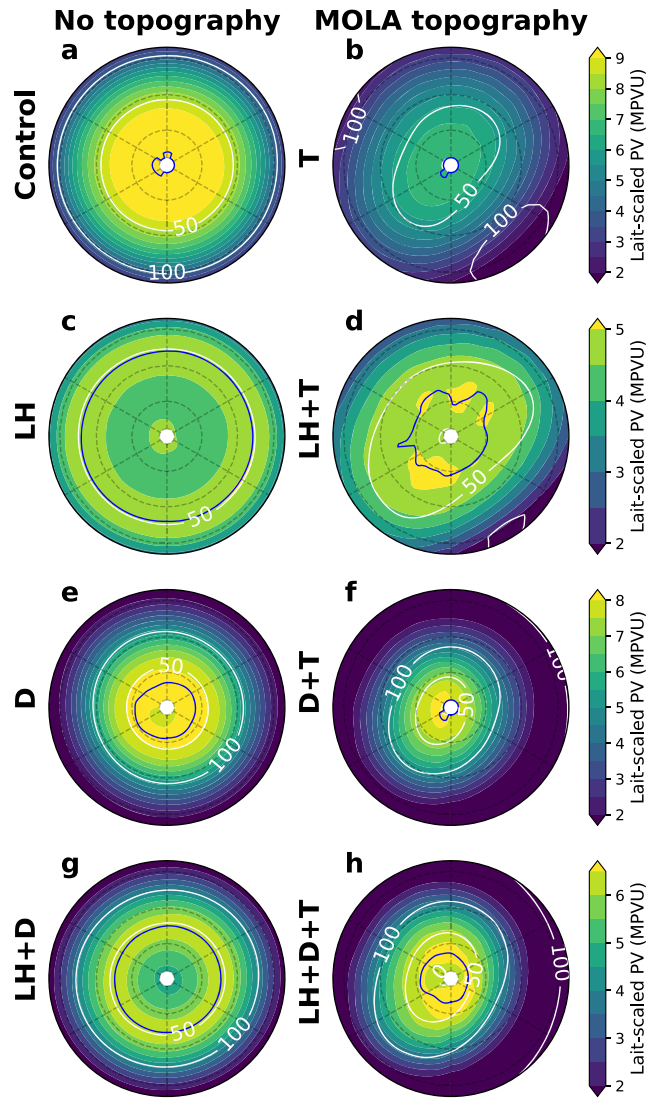

**Figure 6.** Winter-mean ($L_s = 270°$–$300°$) north polar stereographic maps of Lait-scaled PV (shading) and zonal wind (contours) on the 300 K surface from all process-attribution simulations. The solid blue contour shows the latitude of maximum PV. Dashed latitude lines correspond to 60°N, 70°N, and 80°N, with each panel bounded at 55°N. Note that each row has a different color scale.

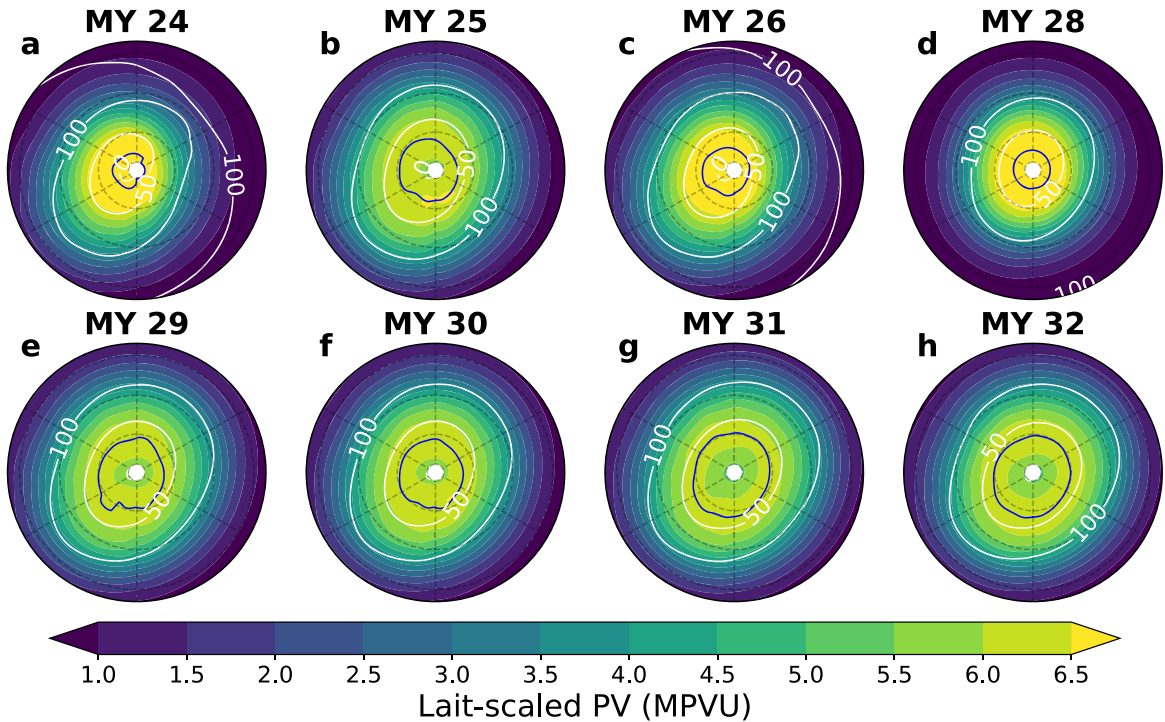

**Figure 7.** Winter-mean ($L_s = 270°–300°$) north polar stereographic maps of Lait-scaled PV (shading) and zonal wind (contours) on the 300 K surface from yearly model simulations. The solid blue contour shows the latitude of maximum PV. See Table 1 for a description of the processes turned on in these simulations. Each year consists of three ensemble members. Dashed latitude lines correspond to 60°N, 70°N, and 80°N, with each panel bounded at 55°N.

28 (Figure A1(d)). Hence, one might conclude that although the dust signal is somewhat weaker in our model than in the reanalysis, dust was indeed a major contributor to the vortex weakening in MY 28. This is in agreement with the nonlinear response of the zonal wind speed to the aerosol heating rate or, equivalently, dust loading (Guzewich et al. 2016).

In the yearly simulations, the morphology of the northern polar vortex also changes somewhat between the TES and MCS eras of observation, as it does in OpenMARS. In the TES era (Figures 7(a)–(c)), the vortex is confined to higher latitudes, while in the MCS era, the annulus is clearer, although it is weaker in terms of its mean PV. This suggests that dust optical depth retrievals from the two instruments may be sufficient to cause some systematic differences in OpenMARS.

Given that Hadley cell outflow has been found to be sufficient to form a PV maximum at 60°N in a full eddy-resolving model (e.g., Scott et al. 2020), changes in the morphology of the polar vortex may well be linked to changes in meridional circulation. For example, Streeter et al. (2021) found that there were longitudinally asymmetric changes in meridional transport into and out of the polar vortices during the MY 34 GDS. To illustrate the difference in circulation between MY 28 and other years in both the reanalysis and model, we now investigate the Eulerian meridional mass stream function, given by

$$\psi(\phi, p) = \frac{2\pi r}{g} \cos\phi \int_0^p \bar{v}(\phi, p)\,dp, \qquad (6)$$

where $r$ is Mars' radius ($3.39 \times 10^6$ m), $g$ is the gravitational acceleration, and $\bar{v}$ is the zonal-mean meridional wind. In both the reanalysis and model, we see the typical Martian cross-equatorial Hadley cell (Figures 8(b) and (e)). In MY 28, this is strengthened, and the descending branch of the overturning cell is pushed further poleward (Figures 8(c) and (f)). Our results in the reanalysis and model are consistent with previous work that has shown that the presence of atmospheric dust strengthens the Hadley circulation, and that a GDS can push the descending branch poleward, particularly at high altitudes (e.g., Guzewich et al. 2016). The model has a persistent anticlockwise cell north of 40°N (Figures 8(d) and (e)). Although this does not appear in the multiannual average for OpenMARS, it is seen in MY 29–32 (not shown), implying differences in meridional circulation during the two observational eras.

### 3.2. Subseasonal Polar Vortex Variability

We now turn to investigating the factors driving shorter-term, subseasonal variability of the northern polar vortex, again drawing on our model simulations and reanalysis data. First, we consider the seasonal evolution of PV, looking at a polar cap average over 60°N–90°N. Figure 9 shows that PV begins to accumulate near the north pole at $L_s \sim 150°$ in both the reanalysis (Figure 9(a)) and model (Figure 9(b)). This occurs regardless of the presence of dust, latent heating, or topography (Figure 9(c)). The PV in the reanalysis appears to evolve differently according to the data assimilated; in both eras, PV initially begins to increase at around $L_s = 150°$, with our model simulations and the reanalysis in good agreement. In the TES era, there is a distinct peak in PV around the winter solstice ($L_s = 270°$) that is not present in the MCS era. In the MCS era, PV reaches its maximum at around $L_s \sim 200°$ and remains at roughly this strength until past $L_s \sim 300°$, when it begins to weaken. The exception to this winter maximum in PV is MY 28, which sees a local minimum in PV at around $L_s \sim 275°$. This corresponds to the weakened PV seen in Figure 2 caused by the GDS. Following the GDS, PV in the vortex recovers once dust loading has reduced. As previously identified, Isca weakly captures this impact in the yearly simulations (Figure 9(b)).

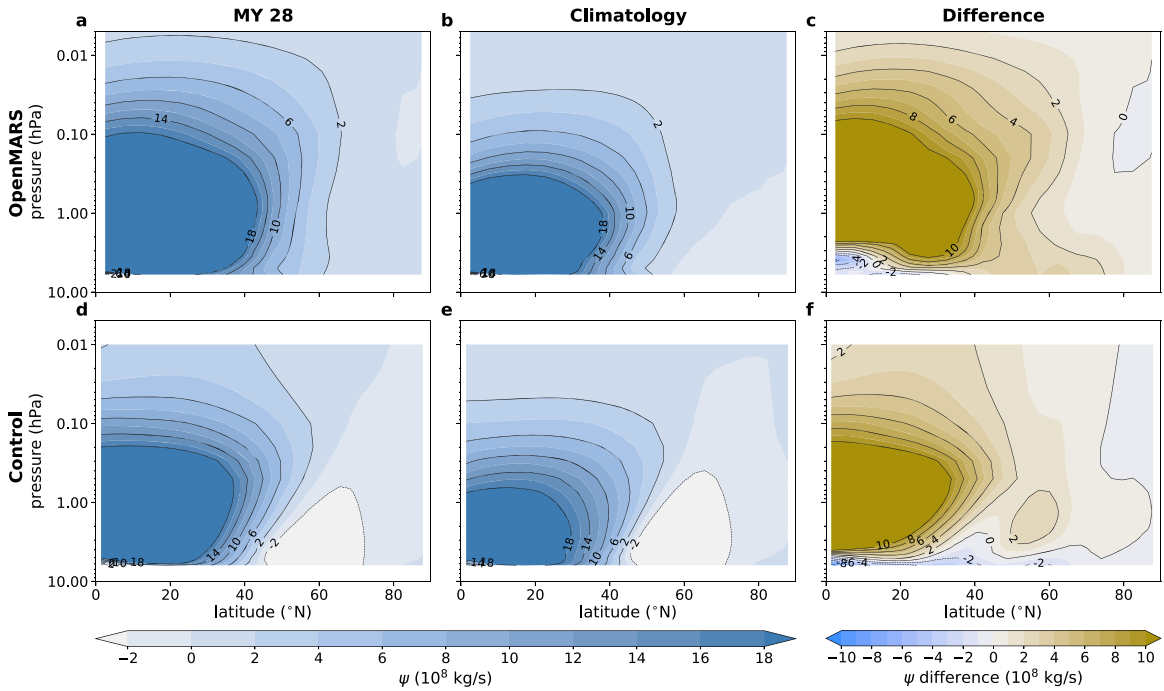

**Figure 8.** Winter ($L_s = 270°–300°$) northern hemisphere Eulerian meridional mass stream function $\psi$ for OpenMARS (top row) and Isca (bottom row). The OpenMARS data are (a) MY 28 and (b) a climatological average over MY 24–26 and 29–32. For Isca, the data shown are (d) the MY 28 yearly simulation and (e) the "latent heat and dust (topography)" simulation. The difference (MY 28 − climatology) is shown for (c) OpenMARS and (f) Isca.

The process-attribution simulations all show a minimum in PV during northern hemisphere summer (Figure 9(c)), as expected. This minimum is deepened by the presence of dust to the point that it agrees well with the reanalysis. However, the winter peak in PV is stronger in the model than in the reanalysis and lasts until the end of the MY. The winter peak in PV is reduced when dust is turned on, particularly during northern hemisphere fall ($L_s \sim 200°$), when the model agrees reasonably well with the reanalysis in all dusty process-attribution simulations. When turned on, both dust and topography always reduce the PV polar cap average throughout the dusty season. Latent heating primarily influences the seasonal evolution of PV between $L_s \sim 200°$ and $360°$, as expected, since this is when the northern hemisphere polar cap is cold enough to allow this process to occur (Figure 1). The effect of latent heating, which is to weaken PV throughout the winter, is predominantly seen in the "control" versus "latent heating" simulations; when dust or topography is turned on, the influence is weaker, suggesting that these are the dominant processes in polar PV evolution.

We also consider the annular nature of the vortex by looking at how the latitude of maximum PV changes over the MY. As previously seen in Figure 3, the maximum in PV can be found well away from the pole in most years (Figure 9(d)), between ~70°N and 80°N over winter. There is more interannual variability in the annularity of the polar vortex in the TES era than in the MCS era of OpenMARS, particularly in fall. In the MCS era, the annulus typically begins to widen just following $L_s \sim 200°$, increasing in width until around $L_s \sim 300°$, where it begins to shrink back toward the pole. In MY 28, the typical widening of the annulus is disrupted, with the maximum in PV found at the pole during the GDS, although the annulus recovers toward the end of the dust storm. This destruction and recovery of the annulus is also captured in our yearly simulations (Figure 9(e)), although we note that in general, the northern polar vortex in our simulations is significantly less annular than the reanalysis. This is, in part, due to the exclusion of any $CO_2$ microphysics and dust as an active tracer in Isca, which means that dust particles cannot act as cloud condensation nuclei for $CO_2$. In spite of this, in our simulations, we do see the larger trend in both PV and the location of its maxima that is visible in TES-era OpenMARS.

Considering the influence of latent heating, dust, and topography on the annularity of the polar vortex (Figure 9(f)) suggests that topography and dust both act to push the maximum in PV further poleward. Indeed, in the "topography" and "dust (topography)" simulations, the maximum is at the pole throughout the lifetime of the polar vortex, indicating an entirely monopolar vortex. The influence of turning on dust is particularly noticeable before $L_s \sim 300°$, with the greatest shift poleward occurring then. This pushes the peak in annularity past $L_s \sim 300°$ in all dusty simulations. As expected, latent heating increases the annularity of the vortex in all cases. The effects of dust and latent heating are of similar magnitude, but the effects of topography are dominant (noting, for example, that in the "latent heating" and "latent heating (topography)" simulations, the PV maximum moves from 70°N to 85°N).

We now look at the internal variability of the northern polar vortex. Scott et al. (2020) investigated the transient nature of the Martian polar vortex by using an eddy-resolving shallow-water model. They considered a quantity called integrated eddy enstrophy, given by

$$Z = \frac{1}{4\pi r^2} \int q'^2 dA, \quad (7)$$

where $q' = q_s - \bar{q}_s$ is the departure from the zonal-mean (scaled) PV and $dA$ is an area integral. Integrated eddy enstrophy is a measure of the flow's zonal asymmetry and is useful in understanding the transience of the flow. They found that a representation of stronger latent heating led to an increased mean and variance of the eddy enstrophy, indicating that both mean eddy activity and the transience of the vortex increase as the vortex becomes more annular and so more unstable. Here we also investigate eddy enstrophy in the northern polar vortex in order to understand how vortex variability develops over the course of a Martian winter and how it is influenced by topography, dust, and latent heat release.

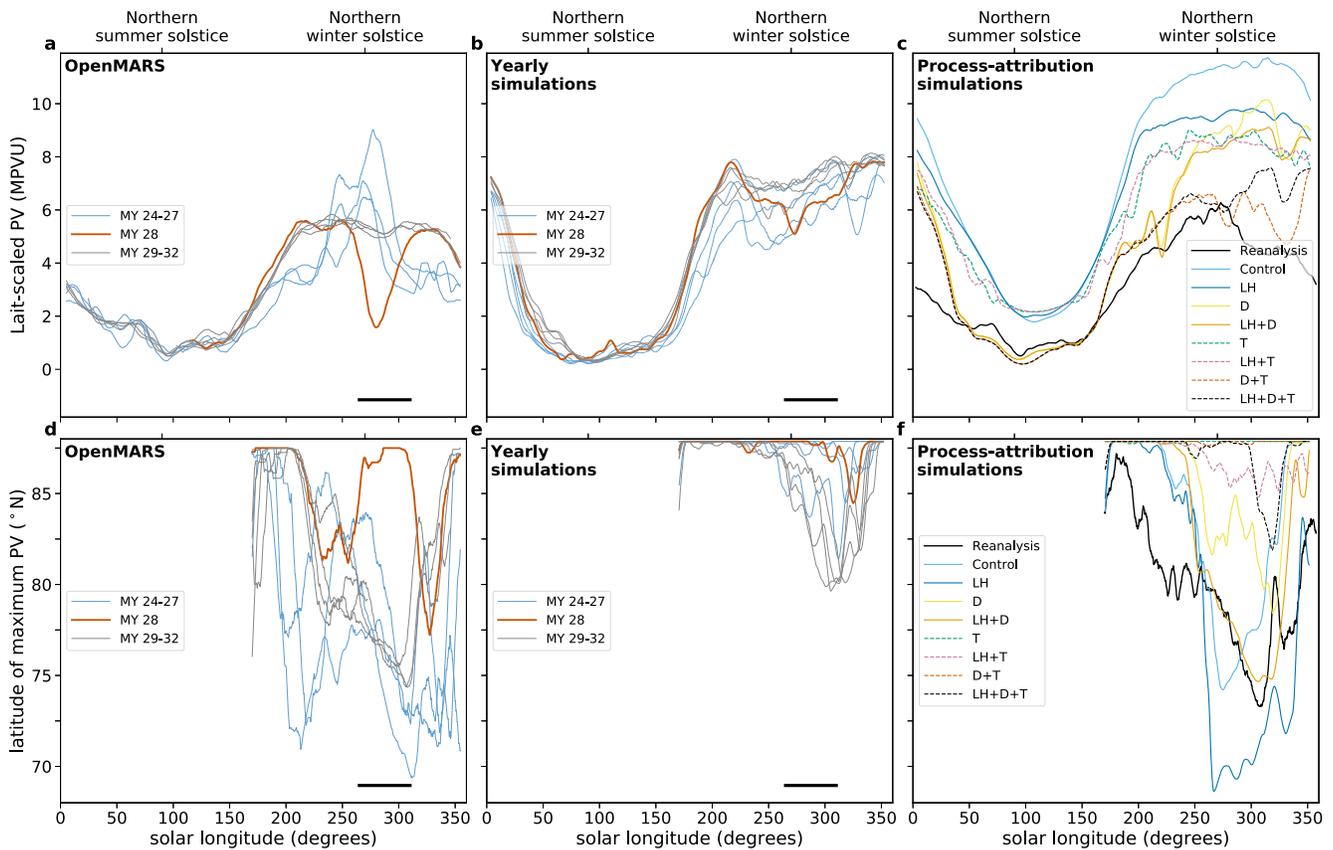

**Figure 9.** Smoothed evolution of mean PV between 60°N and 90°N on the 350 K isentropic surface for (a) OpenMARS, (b) yearly simulations, and (c) process-attribution simulations. Panels (d)–(f) show the corresponding latitude of maximum PV on the 350 K surface. Each simulation is averaged over 3 MY. In panels (a), (b), (d), and (e), lines are colored according to the era of observations (TES, MY 28, and MCS). The solid black line indicates the MY 28 GDS period. In panels (c) and (f), the black line (reanalysis) shows the average of all OpenMARS years, excluding the GDS periods in MYs 25 and 28.

We see in Figure 10 that eddy enstrophy is heavily influenced by dust loading in the Martian atmosphere. In Figures 10(a) and (c), note that MY 28 experiences a significant drop in eddy enstrophy around the winter solstice, at the onset of the GDS. Indeed, in the reanalysis (Figure 10(a)), the eddy enstrophy is almost zero, suggesting that eddy activity within the core of the vortex was minimal. This corroborates the finding that the polar vortex in MY 28 was almost entirely destroyed from a PV point of view.

We also identify here for the first time the interesting double-peaked structure of the eddy enstrophy in the later reanalysis years (Figure 10(a), gray lines), which is not displayed in the earlier ones (blue lines). This pause in eddy enstrophy is reminiscent of the solsticial pause in temperature eddies at low altitudes (Lewis et al. 2016; Mulholland et al. 2016). Intriguingly, this feature is only observed in the OpenMARS reanalysis; there is only a drop in the EMARS eddy enstrophy during MY 28 (see Figure A2(b)). This suggests that the pause in eddy enstrophy is likely a combination of the underlying UK-LMD MGCM used in OpenMARS and a systematic feature of the MCS retrievals, and that the high dust loading in MY 28 does indeed cause a significant drop in eddy enstrophy.

Figures 10(b) and (d) show the effect of topography, latent heating, and dust on eddy enstrophy in the northern polar vortex. The addition of topography to the model increases the eddy enstrophy significantly in the fall ($L_s \sim 200°-270°$). This coincides with the large-amplitude zonal wavenumber 1 waves identified in this season by Wilson et al. (2002), which were noted to have the characteristics of a Rossby wave. The increased variance in the eddy enstrophy on small timescales also indicates that the vortex is more transient when topography is included. In simulations with topography, the inclusion of dust appears to suppress eddy activity around $L_s \sim 0°-75°$ and $150°-225°$ (Figure 10(d)). Without topography, this effect is less substantial and occurs over a shorter period of time, but it is still present. Further analysis is necessary here to investigate the causes of these changes in eddy enstrophy between model configurations, along with the processes that are affected, but this is beyond the scope of this paper.

Overall, we see that it is the combination of dust and topography that allows the model to best capture the shape of the eddy enstrophy evolution seen in the reanalysis, although the peak in eddy enstrophy falls slightly later in the year in the model compared with the reanalysis. Hence, these processes can be seen to be important in capturing realistic vortex variability. Perhaps surprisingly, adding latent heat release to the model does not seem to have a particularly large or consistent effect on the eddy enstrophy. Where latent heat release does cause a drop or increase in eddy enstrophy (e.g., Figure 10(b); $L_s \sim 200°$), we find that this is where there is the largest ensemble spread in our model simulations (not shown).

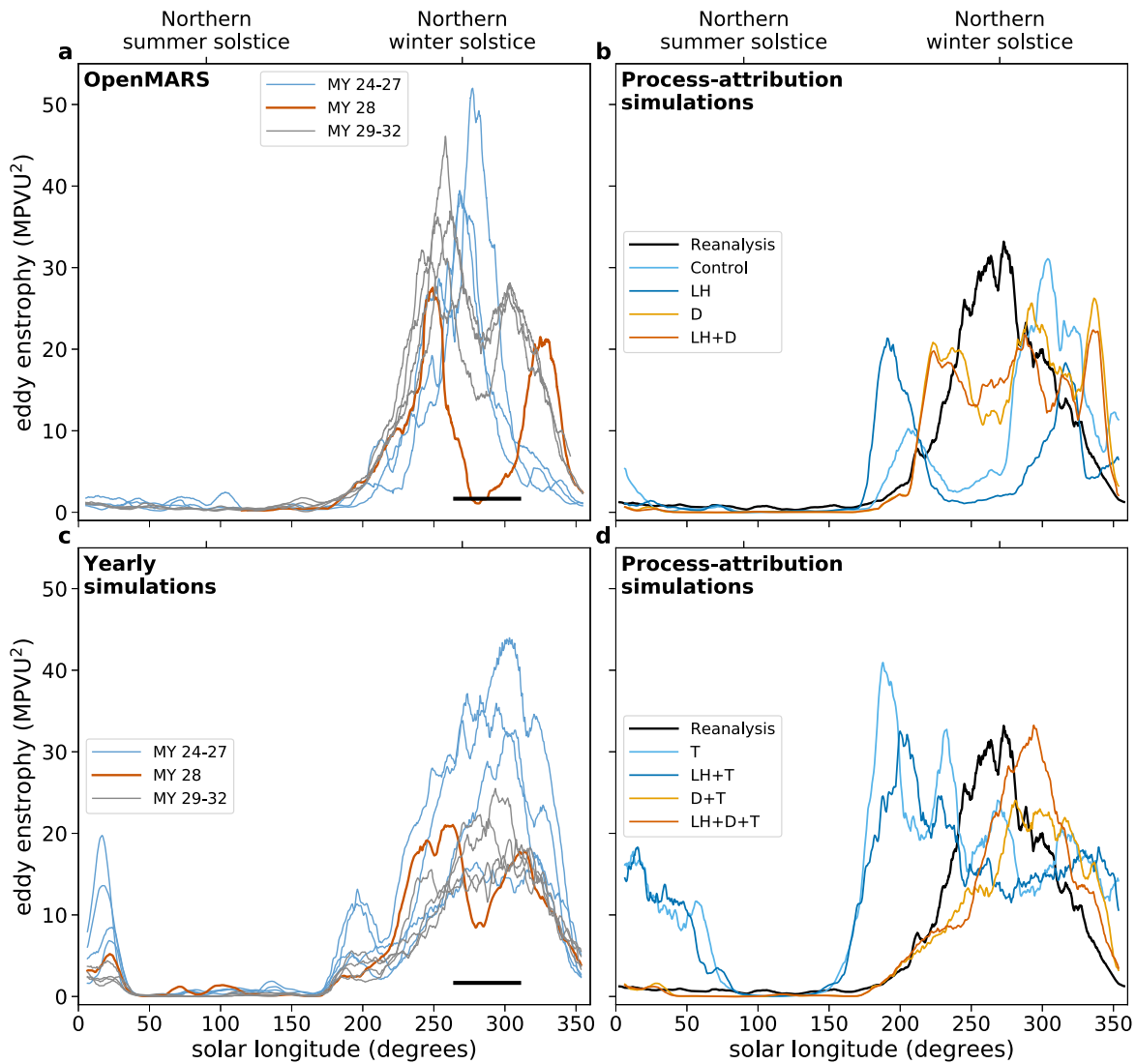

**Figure 10.** Smoothed evolution of eddy enstrophy on the 350 K isentropic surface for (a) OpenMARS and (c) yearly simulations and for process-attribution simulations (b) without topography and (d) with topography. Each simulation is averaged over 3 MY. In panels (a) and (c), the lines are colored according to the era of observations (TES, MY 28, and MCS). The solid black line indicates the MY 28 GDS period. In panels (b) and (d), the black line (reanalysis) shows the average of all OpenMARS years, excluding the GDS periods in MYs 25 and 28.

In general, our simulations show little ensemble spread. In response to the MY 28 dust storm, the polar cap temperature rises, and there is a small increase in zonal wind and decrease in PV on the 350 K surface in all 10 ensemble members. The ensemble spread is greatest in our process-attribution simulations without dust, particularly at times of high eddy activity, such as early fall (not shown). Given the larger ensemble spread at these times, it is difficult to draw conclusions about the influence of latent heat release on mean eddy activity, which is perhaps surprising given the results of Scott et al. (2020). However, we do note consistency with Scott et al. (2020) at the winter peak in eddy enstrophy ($L_s \sim 300°$), which increases in magnitude with the inclusion of latent heating.

When dust is included in the model, it appears to significantly impact the eddy enstrophy evolution, along with topography, far more so than latent heating, noting that the "latent heating" and "latent heating and dust" simulations have broadly the same shape in both panels (b) and (d). This suggests that dust and topography are primarily the driving mechanisms behind eddy enstrophy evolution and hence polar vortex zonal asymmetry and subseasonal variability.

To further investigate the drop in eddy enstrophy noted in MY 28, we also show a temperature proxy for the latent heat released, as in Figure 11. Figure 11(a) shows that in MY 28, following the onset of the GDS, the amount of latent heat released is considerably lower than in all later years during the same time period. This suppression of latent heat release is due to vigorous downwelling at the pole causing higher temperatures within the polar vortex (see Figure 8). This is similarly reflected in the yearly model simulations, where MY 28 once again shows less latent heat release than later years.

We also remark on the apparent difference between the TES and MCS eras in Figure 11(a). It is unclear to us what is causing the particularly small drop below $T_c$ in MY 25–26 and why MY 24 looks different. As there is no evidence of a similar pattern in our simulations (in fact, we see that MY 24 has the least amount of latent heat released in the model TES period), we suggest that this could be caused by temperature retrievals strongly dominating the free-running MGCM in the reanalysis. Differences in the quality control procedures for the TES- and MCS-period dust opacity retrievals mean that only dust opacity retrievals where the surface temperature was greater than 220 K were assimilated into OpenMARS during MY 24–27. On the other hand, both daytime and nighttime dust opacity retrievals from MCS were assimilated (Holmes et al. 2020), allowing the reanalysis to reach these colder temperatures in later years, although this does not explain the difference between MY 24 and MY 25–26. One potential reason for this difference could be the dust optical depths in the reanalysis. Dust loading during the polar winter in the TES era varies from year to year due to the lack of retrievals in this region and the assimilation process (Figure 8 of Holmes et al. 2020). In particular, MY 26 has particularly high dust loading at northern high latitudes, which warms the atmosphere there above the condensation point of $CO_2$.

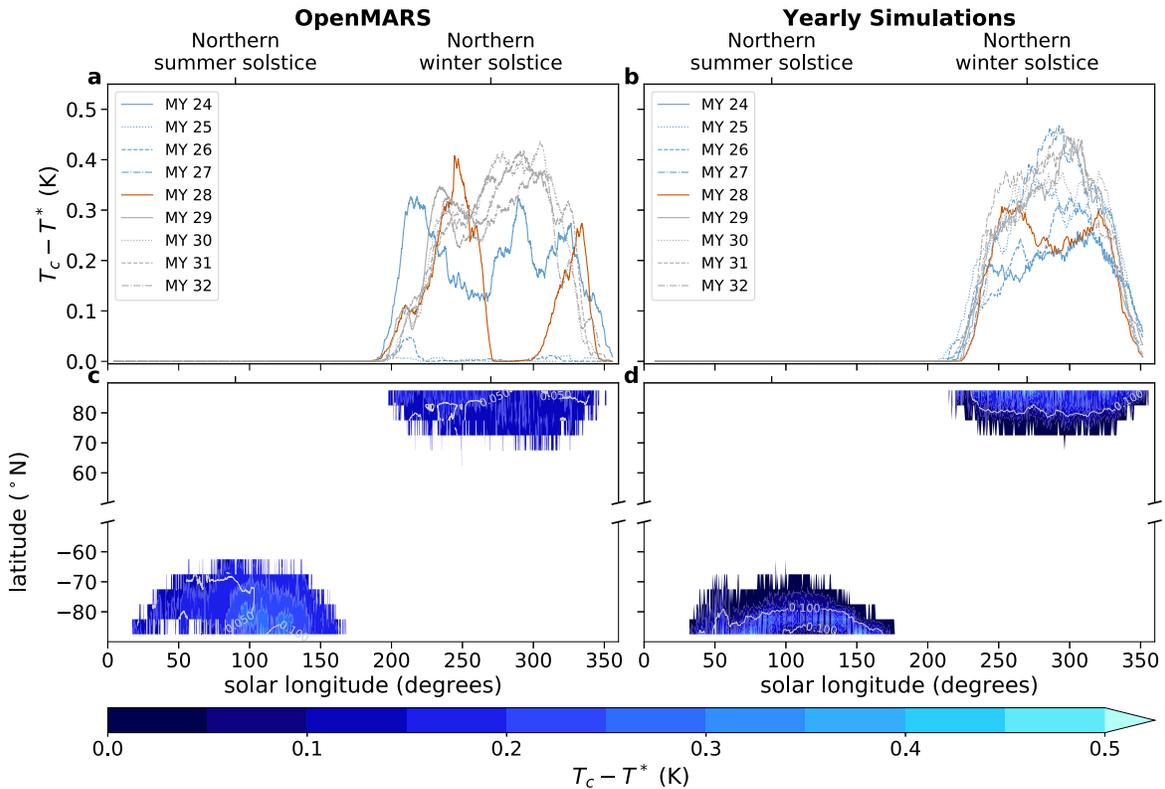

**Figure 11.** Difference $T_c - T'$ (K) for (a) each OpenMARS year and (b) individual yearly simulations on the 2 hPa pressure level. The $T_c$ is given in Equation (3). For panel (a), $T'$ is taken to be the temperature in OpenMARS, and for panel (b), $T'$ is the temperature predicted by the model before the temperature floor at $T_c$ is applied. In panels (a) and (b), we calculate an area-weighted average over 60°N–90°N to show a proxy for the amount of latent heat released. Panels (c) and (d) show the zonally averaged climatology of the evolution of $T_c - T'$ (K) on the 2 hPa pressure level (shading). The climatology is obtained by averaging over MY 24–32 for both (c) OpenMARS and (d) yearly simulations, and the standard deviation (K) is also shown (contours).

Finally, we look at the subseasonal changes in jet strength and overturning circulation during a GDS in Figure 12. During the period of MY 28, the meridional circulation strengthens considerably, the jet weakens, and both the jet latitude and Hadley cell edge shift poleward. We see a difference in the strength of the overturning circulation, defined as the maximum value of $\psi$ at 50 Pa in the northern hemisphere, for the two retrieval eras in Figure 12(a). The mean meridional circulation is somewhat weaker in the TES period than in the MCS period. In the MCS era, the strength of the mean meridional circulation shows little interannual variability, save for MY 28 during the GDS, which strengthens the circulation through aerosol heating. The edge of the Hadley cell, or the latitude at which $\psi$ at 50 Pa first crosses zero north of the equator, is shown for each year of the OpenMARS reanalysis in Figure 12(c). We see that the maximum latitude reached by the Hadley cell edge differs according to the era. During the TES period, the edge of the Hadley cell reaches roughly 75°N, and this maximum occurs just following winter solstice ($L_s \sim 280°$). This is not the case for the MCS period. Instead, there is a local minimum in latitude over the late fall period, and a maximum of around 70°N is reached at $L_s \sim 240°$.

Jet strength is remarkably consistent on the pressure level shown across all years of OpenMARS (Figure 12(b)). The largest deviation during the reanalysis occurs during the MY 28 GDS, when the jet strength is weakened by around 50 m s⁻¹. Figure 12(d) shows the latitude of maximum zonal-mean zonal wind at 50 Pa, the location of the jet core at that altitude. We see that during the MCS era, the jet latitude can be up to 5° poleward of the jet latitude in the TES era throughout fall and winter. Not shown here are the seasonality of the Hadley circulation and polar jet in our model simulations. We find that dust is the primary controlling factor for the strength of the jet and Hadley circulation (strengthening both features). The effects of latent heat release and topography are of relatively minor importance, both acting to weaken the jet slightly.

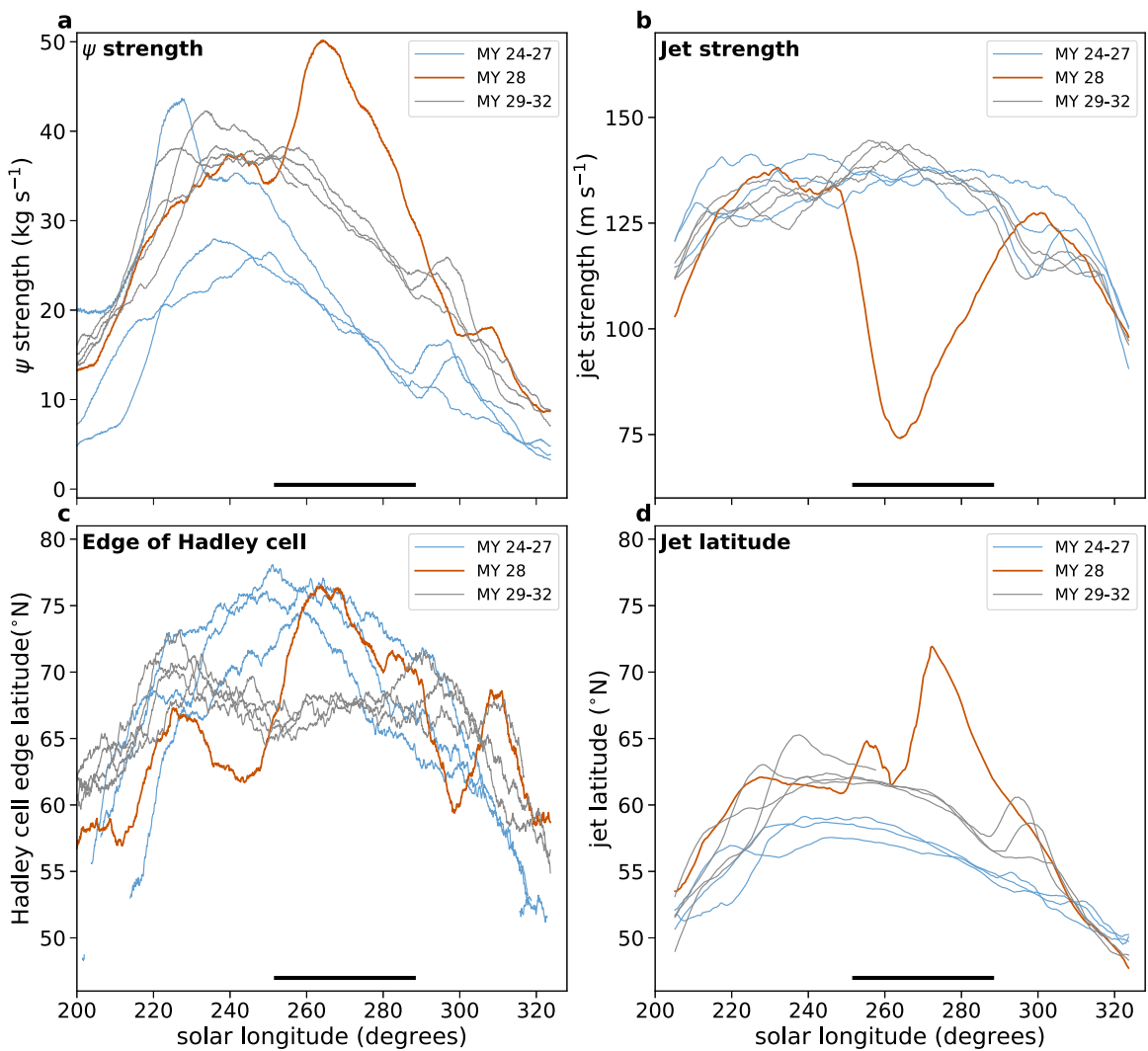

**Figure 12.** Smoothed evolution on the 50 Pa surface of (a) the strength of the overturning stream function and (b) jet strength. The corresponding latitudes of (c) the edge of the Hadley cell and (d) the jet core are shown. We define the jet latitude and strength and the edge and strength of the Hadley cell in the main text. Data are from OpenMARS. Lines are colored according to the era of the data (TES, MY 28, and MCS). The solid black line indicates the MY 28 GDS period.

## 4. Summary

Building on previous work by Guzewich et al. (2016) and Toigo et al. (2017), in this study, we have explored the structure, interannual variability, and subseasonal transience of the northern Martian polar vortex and how they are influenced by an interplay of topography, latent heat release, and dust loading. In agreement with Toigo et al. (2017), we have shown that latent heat release plays a vital role in the annular PV structure of the polar vortex. We showed that the northern polar vortex is heavily affected by the atmospheric dust loading in the reanalysis. In particular, during the GDS of MY 28, there was total destruction of polar PV, and zonal winds were dramatically reduced. During this time, eddy activity within the vortex was also significantly reduced, the Hadley circulation was strengthened, and its edge moved further poleward. We have confirmed the idea that the vertical PV structure of the northern polar vortex in the current reanalyses is well constrained by the observations assimilated into the model, as proposed by Waugh et al. (2016). In the upper levels of the Martian atmosphere, differences correspond to whether there are observations available (during the MCS era) or not (during the TES era). Alongside differences in the mean state of the northern polar vortex, our analysis has also shown significant differences in the subseasonal transience of the northern polar vortex according to which era of the reanalysis is investigated.

We found that seasonal evolution of PV occurs differently in the two periods; there is a winter solstice peak in PV during the TES era that is not seen in the MCS era. Similarly, eddy activity weakens around the winter solstice in MCS-era OpenMARS but not in TES-era. Finally, we find that the edge of the Hadley cell lies further north in the TES period than the MCS but that the jet core lies further north in the MCS period. The meridional overturning circulation is also stronger in the MCS period compared to the TES period.

From our process-attribution simulations, we found that the combination of dust and topography is responsible for the evolution of the zonal asymmetry of the vortex (as measured by eddy enstrophy) throughout the MY, which peaks in the northern high latitudes in winter. Indeed, it is only with the inclusion of dust that we can properly capture the mean state (in particular, the temperatures and winds) and variability of the northern polar vortex in the model. The release of latent heat from carbon dioxide condensation plays a relatively minor role in capturing the magnitude of of eddy activity during northern winter. When there is no dust represented in our model, eddy activity varies significantly from the reanalysis, particularly during northern fall. Further investigation is needed to understand the processes leading to these differences.

It is our hope that the Martian configuration of Isca developed for this work, which significantly extends that of Thomson & Vallis (2019a) by including representations of dust and latent heating, will be useful in future research regarding Martian atmospheric dynamics. While more idealized than some other MGCMs, it has the advantage of being highly flexible and allows users to easily isolate the effects of different parameters and physical processes on the atmospheric circulation. The model allows easy-to-configure representations of latent heat release, dust, and topography, all of which may be adapted to best suit the user's research interests.

Future research may further investigate how different degrees of latent heating affect the annulus within our flexible global climate model by changing the threshold temperature at which carbon dioxide condenses, in a manner similar to Scott et al. (2020). Within this study, we have used a zonally symmetric dust distribution in our simulations. Given that many regional dust storms on Mars can have significant dust loading but do not encircle all longitudes, further research might also address the effects of longitudinally asymmetric dust storms on the Martian polar vortices. We have also focused here on the northern hemisphere polar vortex, but there is plenty of scope to investigate processes affecting the southern hemisphere vortex. Guzewich et al. (2016) showed that the southern vortex was less strongly affected by winter dust loading than the northern, but given the topographical and dust loading hemispheric asymmetries on Mars, such processes may play different roles in the southern polar vortex.

We would like to thank the authors of the OpenMARS and EMARS reanalyses for making their data publicly available.

OpenMARS data can be found in Holmes et al. (2019). EMARS data are available at https://doi.org/10.18113/D3W375. The data derived from OpenMARS, EMARS, and Isca simulations and used to plot the figures within this paper are available from the University of Bristol data repository: doi:10.5523/bris.3i92ii47fndkv2scy9jgrteg0x. We would also like to thank the creators of the MCD for freely distributing the database. The MCD dust products used within this study are available at http://www-Mars.lmd.jussieu.fr/Mars/info_web/index.html; see Madeleine et al. (2011) for details. E.B. is funded by an NERC GW4+ Doctoral Training Partnership studentship from the Natural Environmental Research Council (NE/S007504/1). D.M. is funded under an NERC research fellowship (NE/N014057/1). Finally, we also thank the editor and two anonymous reviewers for their invaluable comments.

*Software:* Isca (Vallis et al. 2018), xarray (Hoyer & Hamman 2017), cartopy (Met Office 2010–2015), MetPy (May et al. 2008–2020), Windspharm (Dawson 2016).

## Appendix
## Supplementary Figures

We present here additional figures intended to supplement the main body of work. Figure A1 shows yearly simulations run with a higher dust loading (a surface mass mixing ratio double that of other simulations; see Table 1 for details). This leads to higher temperatures and stronger winds than those seen in the reanalysis. Significantly, however, we see that PV during the MY 28 dust storm is reduced in these model simulations (Figure A1(d)). Winds are also weakened, in agreement with the nonlinear response to dust loading found in Guzewich et al. (2016).

Figure A2 shows the seasonal evolution of the polar cap PV, the location of the maximum PV, and eddy enstrophy, as calculated from the EMARS reanalysis. Both PV and eddy activity are more consistent between the TES and MCS eras in EMARS than in OpenMARS. The drop in PV and eddy activity during the MY 28 GDS is seen in this reanalysis, as well as in OpenMARS, although the TES peak in PV and the MCS pause in eddy enstrophy are not present. The vortex annularity appears much different, however; it is significantly more annular in the TES era than the MCS.

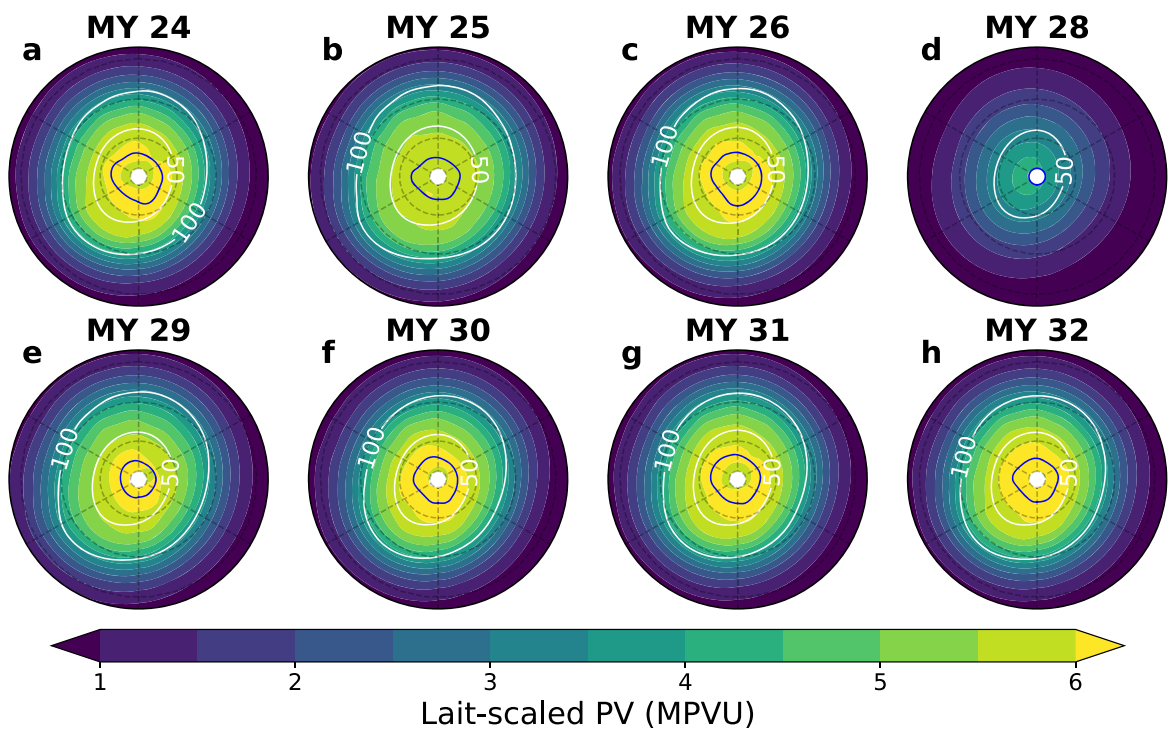

**Figure A1.** Winter-mean ($L_s = 270°$–$300°$) north polar stereographic maps of Lait-scaled PV (shading) and zonal wind (contours) on the 300 K surface from additional "high dust" simulations. The solid blue contour shows the latitude of maximum PV. Dashed latitude lines correspond to 60°N, 70°N, and 80°N, with each panel bounded at 55°N.

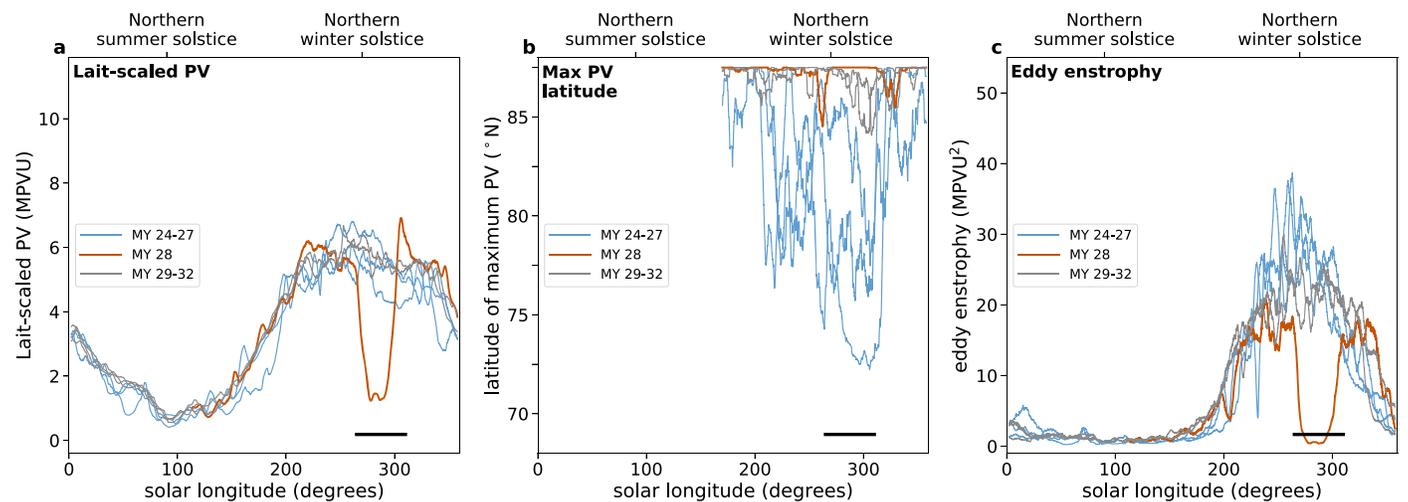

**Figure A2.** Smoothed evolution of (a) polar PV, (b) latitude of maximum PV, and (c) eddy enstrophy on the 350 K isentropic surface for data from the EMARS reanalysis. Polar PV and eddy enstrophy are calculated as in Figures 9 and 10, respectively. Lines are colored according to the era of data (TES, MY 28, and MCS). The solid black line indicates the MY 28 GDS period.

## ORCID iDs

E. R. Ball https://orcid.org/0000-0002-3002-4068
D. M. Mitchell https://orcid.org/0000-0002-0117-3486
W. J. M. Seviour https://orcid.org/0000-0003-1622-0545
S. I. Thomson https://orcid.org/0000-0002-4775-3259
G. K. Vallis https://orcid.org/0000-0002-5971-8995

## References

Ball, E. 2021, BrisClimate/Roles_of_latent_heat_and_dust_on_the_Martian_polar_vortex: Analysis Scripts Second Release, v1.1, Zenodo, doi:10.5281/zenodo.5095055
Banfield, D., Conrath, B., Gierasch, P., Wilson, R., & Smith, M. 2004, Icar, 170, 365
Barnes, J. R., & Haberle, R. M. 1996, JAtS, 53, 3143
Bertrand, T., Wilson, R. J., Kahre, M. A., Urata, R., & Kling, A. 2020, JGRE, 125, e06122
Clancy, R. T., Sandor, B. J., Wolff, M. J., et al. 2000, JGR, 105, 9553
Conrath, B. J. 1975, Icar, 24, 36
Dawson, A. 2016, JORS, 4, 31
Dritschel, D. G., & Polvani, L. M. 1992, JFM, 234, 47
Eckermann, S. D., Ma, J., & Zhu, X. 2011, Icar, 211, 429
Forget, F., Hourdin, F., Fournier, R., et al. 1999, JGR, 104, 24155
Forget, F., Hourdin, F., & Talagrand, O. 1998, Icar, 131, 302
Gooding, J. L. 1986, Icar, 66, 56
Greybush, S. J., Kalnay, E., Wilson, R. J., et al. 2019, GSDJ, 6, 137
Greybush, S. J., Wilson, R. J., Hoffman, R. N., et al. 2012, JGRE, 117, E11008
Guzewich, S. D., Toigo, A., & Waugh, D. 2016, Icar, 278, 100
Hansen, J. E., & Travis, L. D. 1974, SSRv, 16, 527
Harvey, V. L., Randall, C. E., & Hitchman, M. H. 2009, JGRD, 114, D22105
Heavens, N. G., Johnson, M. S., Abdou, W. A., et al. 2014, JGRE, 119, 1748
Hollingsworth, J. L., & Barnes, J. R. 1996, JAtS, 53, 428
Holmes, J., Lewis, S., & Patel, M. 2019, OpenMARS Database (The Open University Collection), doi:10.21954/ou.rd.c.4278950.v1
Holmes, J. A., Lewis, S. R., & Patel, M. R. 2020, P&SS, 188, 104962
Hoskins, B. J., McIntyre, M. E., & Robertson, A. W. 1985, QJRMS, 111, 877
Hoyer, S., & Hamman, J. 2017, JORS, 5, 10
Kass, D. M., Kleinbhl, A., McCleese, D. J., Schofield, J. T., & Smith, M. D. 2016, GeoRL, 43, 6111
Lait, L. R. 1994, JAtS, 51, 1754
Lewis, S. R. 2003, A&G, 44, 4.06
Lewis, S. R., Mulholland, D. P., Read, P. L., et al. 2016, Icar, 264, 456
Madeleine, J.-B., Forget, F., Millour, E., Montabone, L., & Wolff, M. 2011, JGRE, 116, E11010
Madeleine, J.-B., Forget, F., Millour, E., Navarro, T., & Spiga, A. 2012, GeoRL, 39, L23202
Manners, J., Edwards, J. M., Hill, P., & Thelen, J.-C. 2017, SOCRATES Technical Guide: Suite Of Community RAdiative Transfer codes based on Edwards and Slingo, Tech. Rep., UK Met Office, http://homepages.see.leeds.ac.uk/~lecsjed/winscpuse/socrates_userguide.pdf
May, R. M., Arms, S. C., Marsh, P., et al. 2008–2020, MetPy: A Python Package for Meteorological Data, 0.12.1.post2, doi:10.5065/D6WW7G29
Met Office 2010–2015, Cartopy: A Cartographic Python Library with a Matplotlib Interface, Exeter, Devon, https://scitools.org.uk/cartopy
Millour, E., Forget, F., Spiga, A., et al. 2018, Proc. of the Mars Science Workshop "From Mars Express to ExoMars" (Paris: ESA), 68
Mitchell, D. M., Montabone, L., Thomson, S., & Read, P. L. 2015, QJRMS, 141, 550
Montabone, L., Forget, F., Millour, E., et al. 2015, Icar, 251, 65
Montabone, L., Marsh, K., Lewis, S. R., et al. 2014, GSDJ, 1, 129
Montabone, L., Spiga, A., Kass, D. M., et al. 2020, JGRE, 125, e06111
Montmessin, F., Forget, F., Rannou, P., Cabane, M., & Haberle, R. M. 2004, JGRE, 109, E10004
Mulholland, D. P., Lewis, S. R., Read, P. L., Madeleine, J.-B., & Forget, F. 2016, Icar, 264, 465
Pollack, J. B., Colburn, D. S., Flasar, F. M., et al. 1979, JGR, 84, 2929
Read, P. L., Gierasch, P. J., Conrath, B. J., et al. 2007, QJRMS, 132, 1577
Rostami, M., Zeitlin, V., & Montabone, L. 2018, Icar, 314, 376
Scott, R. K., Seviour, W. J. M., & Waugh, D. W. 2020, QJRMS, 146, 2174
Seviour, W. J. M., Waugh, D. W., & Scott, R. K. 2017, JAtS, 74, 1533
Shirley, J. H. 2015, Icar, 251, 128
Shirley, J. H., McKim, R. J., Battalio, J. M., & Kass, D. M. 2020, JGRE, 125, e06077
Smith, D. E., Zuber, M. T., Solomon, S. C., et al. 1999, Sci, 284, 1495
Streeter, P. M., Lewis, S. R., Patel, M. R., et al. 2021, JGRE, 126, e06774
Thomson, S. I., & Vallis, G. K. 2019a, Atmos, 10, 803
Thomson, S. I., & Vallis, G. K. 2019b, QJRMS, 145, 2627
Tillman, J. E. 1988, JGR, 93, 9433
Toigo, A. D., Waugh, D. W., & Guzewich, S. D. 2017, GeoRL, 44, 71
Vallis, G. K., Colyer, G., Geen, R., et al. 2018, GMD, 11, 843
Waugh, D., Toigo, A., & Guzewich, S. 2019, Icar, 317, 148
Waugh, D. W., Toigo, A. D., Guzewich, S. D., et al. 2016, JGRE, 121, 1770
Way, M. J., Aleinov, I., Amundsen, D. S., et al. 2017, ApJS, 231, 12
Wilson, R., Banfield, D., Conrath, B., & Smith, M. 2002, GeoRL, 29, 1684
Wolff, M. J., Smith, M. D., Clancy, R. T., et al. 2006, JGRE, 111, E12S17
Wolff, M. J., Smith, M. D., Clancy, R. T., et al. 2009, JGRE, 114, E00D04
Wolkenberg, P., Giuranna, M., Smith, M., Grassi, D., & Amoroso, M. 2020, JGRE, 125, e06104